\documentclass[aps,pr,twocolumn,groupedaddress,nofootinbib]{revtex4-1}

\usepackage{latexsym}
\usepackage{amsthm}
\usepackage{amsmath}
\usepackage{graphicx}
\usepackage{mathtools}
\usepackage{slashed}
\usepackage{amssymb}
\usepackage{indentfirst}
\usepackage{subfigure}
\usepackage{color}
\usepackage{epsf}
\usepackage[colorlinks=False]{hyperref}

\newcommand*\diff{\mathop{}\!\mathrm{d}}
\newcommand{\nn}{\nonumber}

\newcommand{\be}{\begin{eqnarray}}
\newcommand{\ee}{\end{eqnarray}}
\newcommand{\ma}{\mathrm}
\newcommand{\ml}{\mathcal}
\newcommand{\bs}{\boldsymbol}
\newcommand{\Tr}{\mathrm{Tr}}

\begin{document}

\title{Approach to equilibrium of quarkonium in quark-gluon plasma}

\author{Xiaojun Yao}
\email{xiaojun.yao@duke.edu}
\author{Berndt M\"uller}
\affiliation{Department of Physics, Duke University, Durham, NC 27708, USA}

\date{\today}

\begin{abstract}
We calculate the dissociation and recombination rates of $\Upsilon(1S)$ in quark-gluon plasma by using potential non-relativistic QCD. We then study the dynamical in-medium evolution of the $b\bar{b}-\Upsilon$ system in a periodic box via the Boltzmann equation and explore how the system reaches equilibrium. We find that interactions between the free heavy quarks and the medium
are necessary for the system to reach equilibrium. We find that the angular distribution of $\Upsilon(1S)$ probes the stages at which recombination occurs. Finally, we study the system under a longitudinal expansion and show that different initial conditions evolve to distinct final ratios of hidden and open $b$ flavors. We argue that experimental measurements of the ratio could address open questions in the quarkonium production in heavy ion collisions. 
\end{abstract}

\maketitle

Heavy quarkonium is used as a probe of quark-gluon plasma (QGP) produced in relativistic heavy ion collisions. In their pioneering work, Matsui and Satz~\cite{Matsui:1986dk} argued that Debye screening of the color attraction between heavy quarks ($Q$) and anti-quarks ($\bar{Q}$) leads to the dissociation of the bound state at sufficiently high temperature. As a result, the quarkonium yield was predicted to be suppressed with respect to the scaled yield measured in proton-proton (pp) collision. This ratio is defined as the nuclear modification factor $R_{AA}$. 

The static screening mechanism proposed in the original work is obscured by several factors: nuclear modification of initial production, dynamical screening (dissociation caused by scattering), recombination and feed-down contribution after hadronization. Recent progress in understanding the full complexity of quarkonium suppression includes the calculation of the imaginary potential of $Q\bar{Q}$, which is related to the dissociation rate~\cite{Laine:2006ns, Beraudo:2007ky}; the calculation of the viscous (anisotropic) corrections to the real and imaginary parts of the potential~\cite{Dumitru:2007hy,Dumitru:2009fy,Du:2016wdx}; the calculation of the dissociation rate in potential non-relativistic QCD (pNRQCD)~\cite{Brambilla:2008cx, Brambilla:2011sg, Brambilla:2013dpa} and holographic gravity models of QCD~\cite{Noronha:2009da,Barnard:2017tld}; the description of the time evolution in the open quantum system approach~\cite{Young:2010jq, Borghini:2011ms, Akamatsu:2011se, Akamatsu:2014qsa, Blaizot:2015hya, Brambilla:2016wgg, Kajimoto:2017rel, DeBoni:2017ocl}; the use of Langevin equations to study the impact of $Q\bar{Q}$ diffusion~\cite{Petreczky:2016etz} on quarkonia recombinations~\cite{Young:2008he, Young:2009tj, Chen:2016mhl, Zhao:2017yan, Chen:2017duy}; and phenomenological studies that can reproduce the experimentally measured $R_{AA}$ of bottomonia with~\cite{Du:2017qkv} and without recombination~\cite{Krouppa:2015yoa}.

The process of recombination is less well-understood than dissociation. Different studies incorporate recombination in varying ways. In the open quantum system formalism, an initially unbound $Q\bar{Q}$ pair will evolve to have non-vanishing overlap with bound state wave functions due to the action of stochastic potentials. One-dimensional numerical studies in the abelian case have been carried out~\cite{Akamatsu:2011se, Blaizot:2015hya, Kajimoto:2017rel, DeBoni:2017ocl} and the non-abelian case has been developed formally~\cite{Akamatsu:2014qsa}. Other approaches use coalescence models based on Wigner functions~\cite{Young:2008he, Chen:2017duy} or invoke detailed balance to model the recombination rate as the dissociation rate times equilibrium quarkonium fraction~\cite{Du:2017qkv}, which is only true near chemical and thermal equilibrium. Here we aim to present a calculation that incorporates dissociation and recombination in a consistent way and without assuming the quarkonium distribution is close to equilibrium.

Recombination is believed to be less important for bottomonium than charmonium because fewer $b$-quarks are produced in the collision. This would be consistent with experimental measurements if the dissociation is the dominant in-medium process, which is based on the assumption of small suppression of initial quarkonium production. However, we only know the nuclear modification on parton distribution functions is small~\cite{Ferreiro:2008wc}, which is not sufficient for the assumption. Quarkonium production in pp collisions factorizes into short-distance production of heavy quarks and long-distance coalescence into quarkonium~\cite{Bodwin:1994jh}. It is unlikely that the long-distance physics completes before the formation of QGP in heavy ion collisions. Therefore the fraction of quarkonia formed in the initial stage is not well-determined.

Furthermore, if the initial QGP temperature is higher than the melting temperature above which certain quarkonium state cannot exist due to static screening (which has been studied from the temperature dependence of the binding energy~\cite{Karsch:1987pv} or spectral functions~\cite{Mocsy:2007jz} in potential models), correlated $Q\bar{Q}$ pairs will remain unbound when entering QGP and may (re)combine later. This type of process is also defined as recombination throughout this paper. In short, recombination of correlated and uncorrelated heavy quark anti-quark pairs may be more important than originally thought.

To illustrate these points, we propose a dynamical in-medium transport model based on Boltzmann evolution with dissociation and recombination. This approach allows us to explore how experiments can help address various unknown aspects of quarkonium production mechanisms in heavy ion collisions. Here we consider the $\Upsilon(1S)$ as an example. The physical picture is as follows: if the local temperature is higher than the melting temperature, $\Upsilon(1S)$ dissociates and locally only unbound $b\bar{b}$ exist. However, if the local temperature is lower, on one hand, $\Upsilon(1S)$ can exist and propagate in the medium and may be dissociated by scattering with medium gluons and light quarks; on the other hand, unbound $b$ and $\bar{b}$ propagate and diffuse, and at any time, they may recombine into $\Upsilon(1S)$ by scattering with medium constitutes, if they are sufficiently close to each other and their relative momentum favors recombination.

We calculate the dissociation and recombination rates to lowest order in pNRQCD~\cite{Brambilla:1999xf,Brambilla:2004jw}. The effective theory can be derived from QCD under the hierarchy of scales $M \gg Mv \gg Mv^2, T, m_D$ where $M=4.65$ GeV is the $b$-quark mass, $v\sim0.3$ is the relative velocity of $b\bar{b}$ inside $\Upsilon(1S)$, $T$ is the temperature, and $m_D$ is the Debye screening mass. The pNRQCD Lagrangian is given by
\begin{widetext}
\be
\label{eq:lagr}
\ml{L}_\ma{pNRQCD} = \int d^3r \Tr\Big(  \ma{S}^{\dagger}(i\partial_0-H_s)\ma{S} +\ma{O}^{\dagger}( iD_0-H_o )\ma{O} 
+V_A( \ma{O}^{\dagger}\bs r \cdot g{\bs E} \ma{S} + \ma{h.c.})  + \frac{V_B}{2}\ma{O}^{\dagger}\{ \bs r\cdot g\bs E, \ma{O}  \} +\cdots \Big)\, ,
\ee
\end{widetext}
where ${\bs E}$ represents the color electric gauge field. The Lagrangian of gluon and light quark is just QCD with momenta $\lesssim Mv$. The pNRQCD is a systematic expansion in $v$ or $1/M$ (NR expansion) and $r$,  the relative distance between $b\bar{b}$ (multipole expansion). The degrees of freedom are the color singlet $\ma{S}(\bs R, \bs r, t)$ and color octet $\ma{O}(\bs R, \bs r, t)$ states where $\bs R$ denotes the center-of-mass (c.m.) position and $\bs r$ the relative coordinate. The color singlet and octet Hamiltonians are expanded in powers of $1/M$:
\be
H_{s,o} = \frac{\bs P_\ma{cm}^2}{4M} + \frac{\bs p_\ma{rel}^2}{M} + V_{s,o}^{(0)} + \frac{V_{s,o}^{(1)}}{M} + \frac{V_{s,o}^{(2)}}{M^2} + \cdots\,.
\ee
By the virial theorem, $\bs p_\ma{rel}^2/M \sim V_{s,o}^{(0)}\sim Mv^2$. Higher-order terms of potentials including relativistic corrections, spin-orbital and spin-spin interactions are further suppressed by extra powers of $v$. The c.m. kinetic energy is also suppressed because momenta $\sim Mv$ have been integrated out in the construction so ${\bs P}_\ma{cm}\ll Mv$. We only work to order $Mv^2$ because $v$ is small and also heavy ion experiments do not resolve hyperfine structures currently.
In the following, we only consider temperatures at which the $\Upsilon(1S)$ exists ($T_C < T < 2.5T_C$, $T_C=155$ MeV). In this domain the confining potential is flattened and potentials can be approximated by Coulomb interactions
\be
V_{s}^{(0)} = -C_F\frac{\alpha_s}{r},\ \ \ \ \ \ \ V_{o}^{(0)} = \frac{1}{2N_c}\frac{\alpha_s}{r}\,.
\ee
The singlet-octet and octet-octet vertices are color dipole interactions with $V_A=V_B=1$.

\begin{figure}
\centering
\vspace{0.1in}
\includegraphics[width=0.9\linewidth]{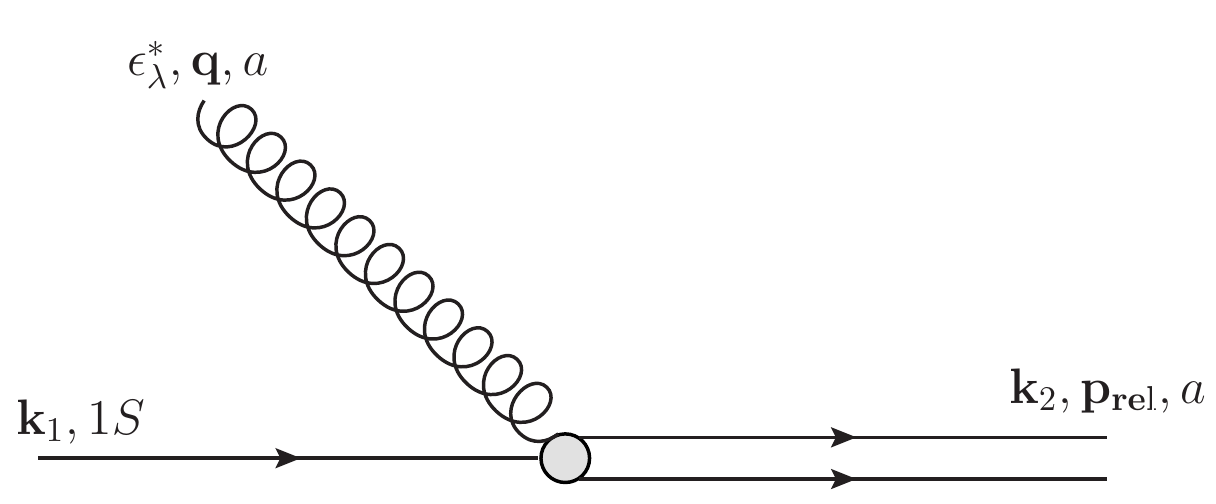}
\caption{Transition between $\Upsilon(1S)$ and $b\bar{b}$ octet by absorbing or emitting an on-shell gluon. Single line indicates quarkonium while double lines represent unbound octet.}\label{fig:s_o}
\end{figure}

The Feynman diagram of the transition between the singlet $\Upsilon(1S)$ and the unbound $b\bar{b}$ octet via absorption or emission of a gluon is shown in Fig.~\ref{fig:s_o}. For simplicity, we here consider only the interaction with on-shell gluons in the QGP. Transitions mediated by virtual gluons (inelastic scattering with medium constitutes) are at next order in $\alpha_s$ and can be easily included within our formalism. But we are also aware that when $m_D\gg E_{1S}$, the inelastic scattering dominates~\cite{Brambilla:2013dpa}. Our future full calculations will include both. The scattering amplitude is given by
\be
\label{eqn:amp}
\ml{T}^a &=& (2\pi)^4 \delta^3({\bs q} + {\bs k}_1 - {\bs k}_2) \delta(\Delta E) \ml{M}^a
\\\nn
\ml{M}^a &=& -ig\sqrt{\frac{T_F}{N_c}}q\langle \psi_{1S} | {\bs \epsilon}_{\lambda}^*\cdot {\bs r} | \Psi_{\bs p_\ma{rel}} \rangle
\\\nn
\Delta E &=& q+\frac{k_1^2}{4M}+E_{1S}-\frac{k_2^2}{4M}-\frac{p_\ma{rel}^2}{M} \,.
\ee
where $T_F=1/2$, $|\psi_{1S}\rangle$ is the hydrogen-like $1S$ wave function for $\Upsilon(1S)$ and $| \Psi_{\bs p_\ma{rel}} \rangle$ is the Coulomb wave function for unbound $b\bar{b}$ octet. The $1S$ binding energy is given by $E_{1S} = - \alpha_s^2C_F^2M/4$ and the gluon energy is $q = |{\bs q}|$. Throughout our paper we set $\alpha_s = 0.3$. Here ${\bs k}_{1,2}$ are c.m. momenta and their associated kinetic energies will be neglected according to the power counting.

The set of Boltzmann equations for the $b$, $\bar{b}$ and $\Upsilon(1S)$ distribution functions $f_i({\bs x}, {\bs p}, t)$ is given by
\be
\label{eq:LBE}
(\frac{\partial}{\partial t} + \dot{{\bs x}}\cdot \nabla_{\bs x})f_i({\bs x}, {\bs p}, t) &=& \ml{C}_i  - \ml{C}_+ + \ml{C}_-\\\nn
(\frac{\partial}{\partial t} + \dot{{\bs x}}\cdot \nabla_{\bs x})f_\Upsilon({\bs x}, {\bs p}, t) &=& \ml{C}_+ - \ml{C}_-\,,
\ee
where $i=b$ or $\bar{b}$. For $b$ and $\bar{b}$ quarks, the collision term $\ml{C}_i$ describes their scattering with thermal constituents of QGP. This process has been described either as diffusion in the framework of the Langevin equation~\cite{Moore:2004tg,Xu:2017hgt} or as two-body scattering in the framework of the linearized Boltzmann equation~\cite{Gossiaux:2008jv,Gossiaux:2009mk,Uphoff:2014hza}. Here we use, for simplicity, the relaxation-time approximation with $\ml{C}_i = -\Gamma_r(f_i-f_i^\ma{eq})$. The relaxation rate is assumed to be $\Gamma_r=T^2/M$~\cite{Moore:2004tg} and $f_i^\ma{eq}$ is the relativistic Boltzmann distribution. For $\Upsilon(1S)$, the gain term $\ml{C}_+$ is from recombination by gluon emission and the loss term $\ml{C}_-$ is from dissociation by gluon absorption\footnote{In the original print, a factor $(2\pi)^3\delta^3({\bs k}_1 - {\bs p})$ was missed inside the integral of expression (\ref{eqn:reco}).}:
\begin{widetext}
\be
\label{eqn:reco}
\nn
\ml{C}_+ &=& \int\frac{\diff^3p_1}{(2\pi)^3}\frac{\diff^3p_2}{(2\pi)^3}\frac{\diff^3k_1}{(2\pi)^3}\frac{\diff^3q}{2q(2\pi)^3}(1+n_B^{(q)})
\frac{8}{9}f_b({\bs x}, {\bs p}_1, t)f_{\bar{b}}({\bs x}, {\bs p}_2, t)(2\pi)^7\delta^3({\bs q} + {\bs k}_1 - {\bs k}_2) \delta(\Delta E)\delta^3({\bs k}_1-{\bs p})\overline{|\ml{M}^a|^2}
\\
\\
\ml{C}_- &=& \frac{1}{\gamma}\int\frac{\diff^3p_\ma{rel}}{(2\pi)^3}\frac{\diff^3k_2}{(2\pi)^3}\frac{\diff^3q}{2q(2\pi)^3}n_B^{(q)}
(2\pi)^4 \delta^3({\bs q} + {\bs k}_1 - {\bs k}_2) \delta(\Delta E) \overline{|\ml{M}^a|^2}f_{\Upsilon}({\bs x}, {\bs p}, t)
\equiv \Gamma_d({\bs x}, {\bs p}, t)f_{\Upsilon}({\bs x}, {\bs p}, t)\, ,
\ee
\end{widetext}
where the second equation defines the gluo-dissociation rate $\Gamma_d$. The scattering amplitude is calculated in the rest frame of $\Upsilon$ for dissociation and that of $b\bar{b}$ for recombination, where the pNRQCD is valid. The Bose distribution of medium gluons $n_B^{(q)}$ is boosted into the rest frame of $\Upsilon$ or $b\bar{b}$ accordingly. The two frames are not equivalent but since the gluon energy is small compared to $M$ ($T\ll M$), the difference is suppressed by $T/M$. The overline indicates an average over initial-state and sum over final-state quantum numbers (color and spin). The phase space measure is relativistic for gluons and non-relativistic for $b$-quarks, which is consistent with our field definitions.

For dissociation, the rest-frame rate is then boosted back into the medium frame by the factor $\gamma^{-1} = \sqrt{1-v^2}$ where $v$ is the $\Upsilon$ velocity. For recombination, the gamma factor cancels out, as explained further below. The quark momenta ${\bs p}_1$ and ${\bs p}_2$ are related to the relative momentum ${\bs p}_\ma{rel}$ used in the amplitude calculation via $\frac{1}{2}({\bs p}'_1 - {\bs p}'_2 )$ where the primed momenta are in the $b\bar{b}$ rest frame. The factor $8/9$ ensures that only a color octet $b\bar{b}$ pair can form a singlet bound state by emitting a gluon. 

We solve the Boltzmann equations (\ref{eq:LBE}) by stochastic simulation. Here we study the evolution inside a periodic box of QGP with side length $L=10$ fm. The QGP temperature is constant throughout the box but can change with time. A certain number of $b$, $\bar{b}$ and $\Upsilon$ ($N_b$, $N_{\bar{b}}$ and $N_{\Upsilon}$) are initialized by random sampling of their positions and momenta, assuming a given initial momentum distribution. At each time step $\Delta t$, we consider three types of processes:

First, for each $\Upsilon$ with a given velocity, we determine whether it dissociates according to the probability $\Gamma_d \Delta t$. If it dissociates, we sample the incoming gluon momentum in the rest frame of $\Upsilon$ according to the integrand of the rate integral and then calculate the outgoing relative momentum of $b\bar{b}$ by energy-momentum conservation. Finally we boost the momenta of $b$ and $\bar{b}$ back into the medium frame; their positions are set to be the position of $\Upsilon$ before dissociation.

For each $b$-quark with position ${\bs y}_i$ and momentum ${\bs k}_i$ we need to determine the total recombination rate with neighboring $\bar{b}$ quarks with position ${\bs z}_j$ and momenta ${\bs k}_j$. However, the quark and anti-quark distributions in the expression (\ref{eqn:reco}) should be evaluated at the same position, but the product of two delta functions is ill-defined. We introduce a position-dependence of the recombination probability by means of a Gaussian function with a width $\sigma$ chosen to be the $\Upsilon(1S)$ Bohr radius. This ensures that the recombination probability of a widely separated $b\bar{b}$ pair vanishes. The product of local distributions in (\ref{eqn:reco}) is thus replaced with
\begin{widetext}
\be
\label{eq:ff}
f_b({\bs x}, {\bs p}_1, t)f_{\bar{b}}({\bs x}, {\bs p}_2, t) \rightarrow      
\sum_{i,j}  \frac{e^{-({\bs z}_j - {\bs y}_i)^2/2\sigma^2}}{(2\pi\sigma^2)^{3/2}} \delta^3\left({\bs x}-\frac{{\bs y}_i+{\bs z}_j}{2}\right)   
2\theta\left[-({\bs z}_j-{\bs y}_i)\cdot({\bs k}_j-{\bs k}_i)\right]  \delta^3({\bs p}_1-{\bs k}_i) \delta^3({\bs p}_2-{\bs k}_j)\,,
\ee
\end{widetext}
where the sum over $i,j$ runs over all $b,\bar{b}$ contained in the box. The position-dependence (including the choice of $\sigma$) disappears if one averages over many spatial configurations. The theta function assures that only approaching quark pairs can recombine. Thus, a $b\bar{b}$ that has just dissociated cannot recombine until at least one of them scatters once.

For a given $b$-quark, the $\bar{b}$ density is Lorentz boosted into the rest frame of $b\bar{b}$, and when the rate is transformed back into the medium frame, the two $\gamma$ factors cancel. This explains the absence of gamma factor in (\ref{eqn:reco}). If the $b$-quark is found to recombine, a $\bar{b}$-quark is chosen based on the value of recombination probability. We then sample the outgoing gluon and replace the $b\bar{b}$ pair with a $\Upsilon$ whose momentum is determined from energy-momentum conservation. Its position is given by the c.m. position of the quark pair as indicated in (\ref{eq:ff}).

Third, the diffusion of unbound $b$ or $\bar{b}$ quark in the QGP is implemented by re-sampling its momentum from thermal distribution at a probability of $\Gamma_r\Delta t$ in each time step.
We exclude elastic scattering between medium particles and bound $\Upsilon$ because it cannot happen at the order we are working.

\begin{figure}
\centering
\includegraphics[width=1\linewidth]{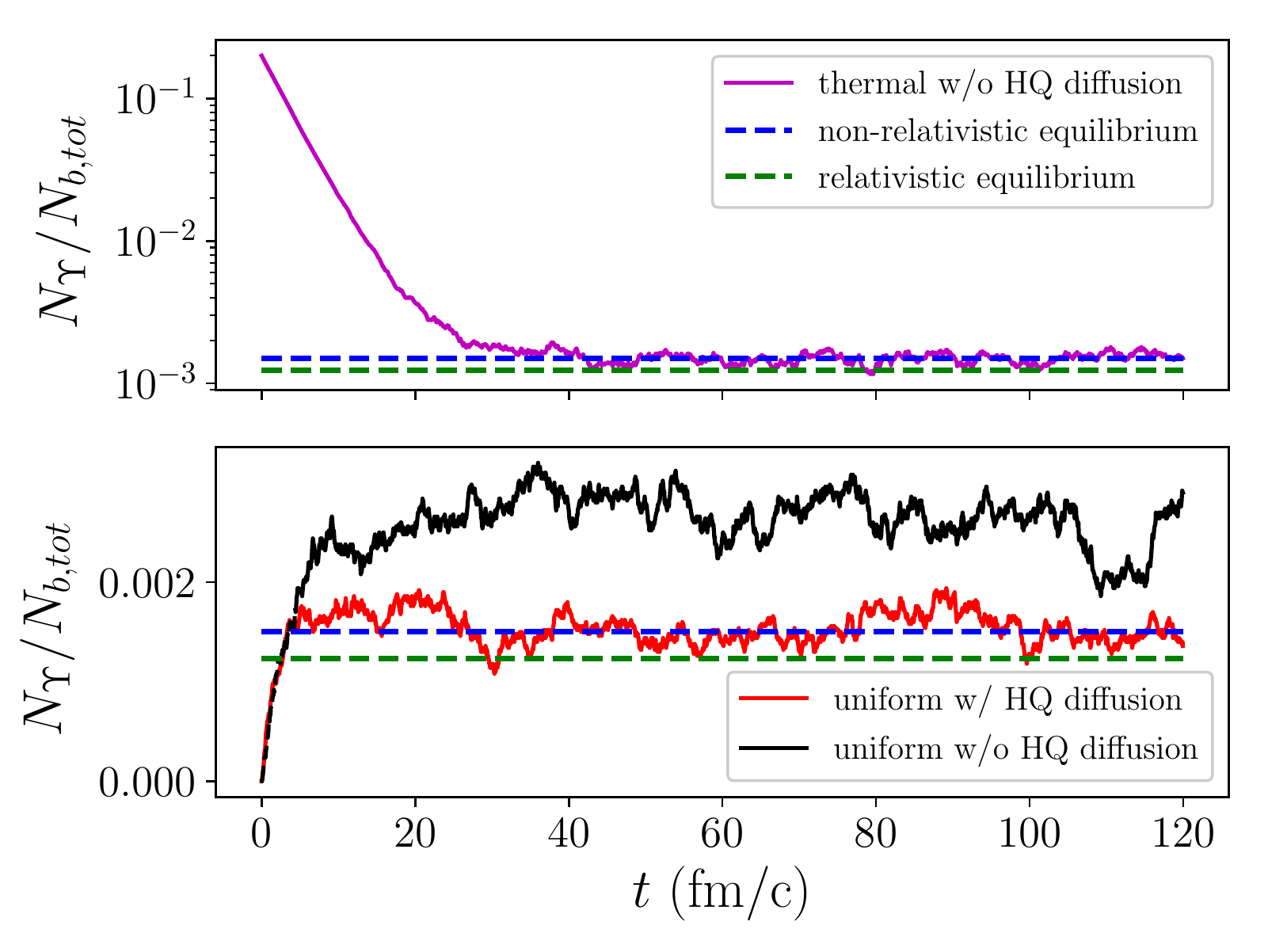}
\caption{(Color online) Simulations at $T=350$ MeV. $N_b=N_{\bar{b}}=40$ and $N_{\Upsilon}=10$ (case 1) with thermal momenta (upper) and $N_b=N_{\bar{b}}=50$ and $N_{\Upsilon}=0$ (case 2, 3) with uniform momenta (lower). The dashed lines indicate the abundance ratio at equilibrium.}\label{fig:det_bal}
\end{figure}

As a first application, we study how the $b\bar{b}$-$\Upsilon$ system reaches chemical equilibrium. We set up the system at a constant temperature $T=350$ MeV with $N_{b,\ma{tot}}=N_b+N_{\Upsilon}=50$ in three different initial conditions: 
\begin{enumerate}
\item $N_b=N_{\bar{b}}=40$, $N_{\Upsilon} = 10$; the initial momenta of all particles are sampled from thermal Boltzmann distributions with relativistic dispersion relation;
\item $N_b=N_{\bar{b}}=50$, $N_{\Upsilon} = 0$; the initial momentum components of all particles are sampled uniformly in the range $-1\ \ma{GeV} < p_x,p_y,p_z < 1\ \ma{GeV}$ with heavy quark (HQ) diffusion turned off;
\item as case 2, but including HQ diffusion.
\end{enumerate}
The results of $N_{\Upsilon}/N_{b,\ma{tot}}$ are plotted in Fig.~\ref{fig:det_bal}. At equilibrium
\be
N^\ma{eq}_{i}=g_i\ma{Vol}\int\frac{\diff^3p}{(2\pi)^3}\lambda_{i}e^{-E_i(p)/T}\,,
\ee 
with $E_i(p)=\sqrt{M_i^2+p^2}$ relativistically and $M_i+\frac{p^2}{2M_i}$ non-relativistically for $i=b,\bar{b}$ or $\Upsilon$. The degeneracy factors are $g_b=g_{\bar{b}}=6$ and $g_{\Upsilon}=4$ (because hyperfine splitting is not considered here and thus $\eta_b$ and $\Upsilon(1S)$ are degenerate). The fugacities are related by $\lambda_{\Upsilon}=\lambda_{b}\lambda_{\bar{b}}=\lambda_{b}^2$ and solved from $N^\ma{eq}_{b}+N^\ma{eq}_{\Upsilon}=N_{b,\ma{tot}}$. The simulations converge to the NR lines because the rates are calculated in a pNRQCD. If excited states are included, the $\Upsilon(1S)$ equilibrium fraction will decrease but only insignificantly. As the lower part of Fig.~\ref{fig:det_bal} shows, HQ diffusion is necessary for the system to reach equilibrium starting from a non-thermal initial distribution.

\begin{figure}
\centering
\includegraphics[width=1\linewidth]{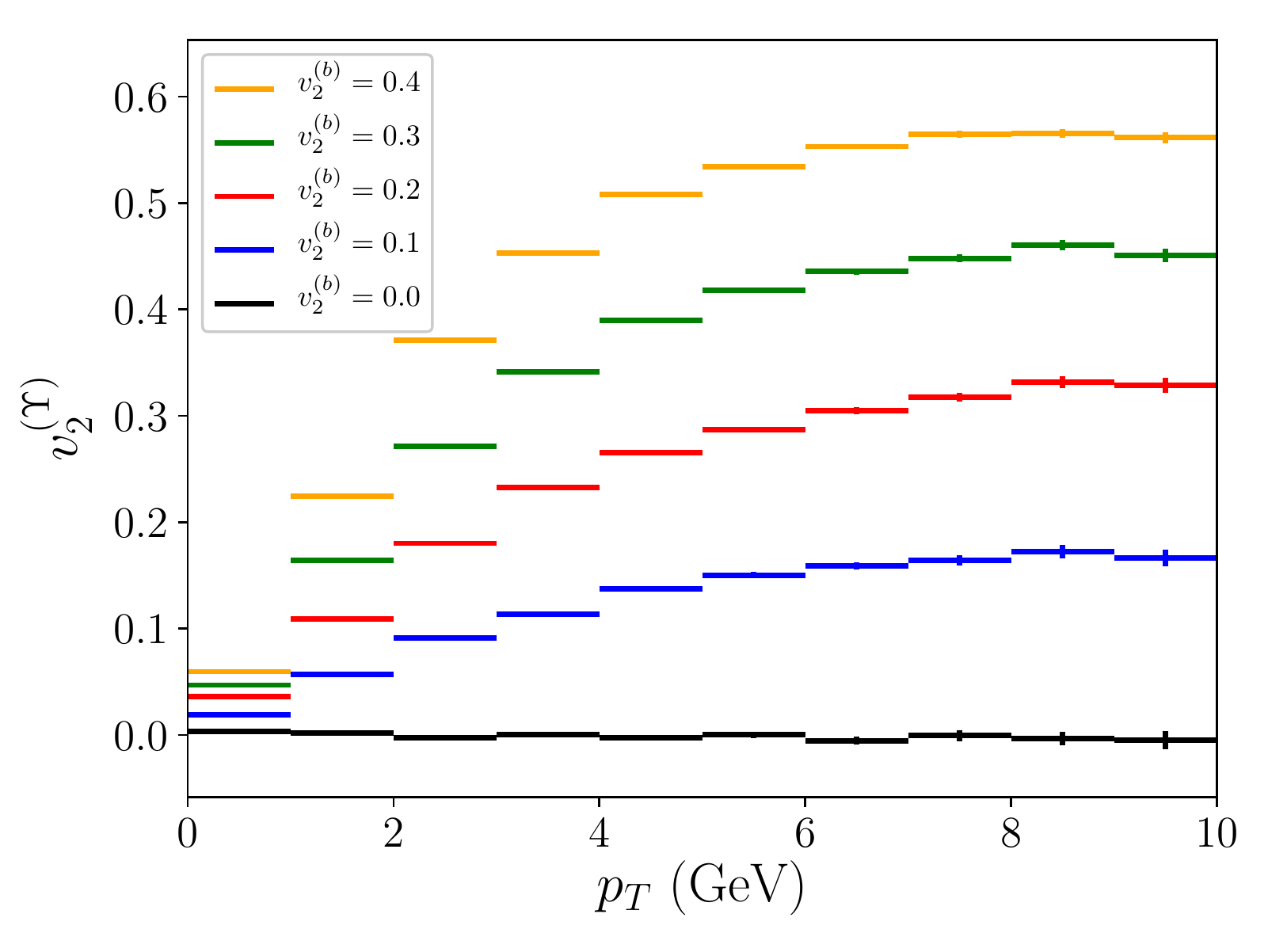}
\caption{(Color online) Angular anisotropy $v_2$ of $\Upsilon$ from recombinations of $b,\bar{b}$ with anisotropic momentum distributions for different values of $v_2^{(b)}$.}\label{fig:upsilon_v2}
\end{figure}

We next study the azimuthal angular anisotropy of $\Upsilon$ produced from recombinations of $b$ and $\bar{b}$ with certain azimuthal momentum distributions simulating elliptic flow of the QGP, which is gradually transmitted to the unbound heavy quarks by diffusion during the QGP phase. Since quarkonia can form at any time below the melting temperature and not necessarily have to wait until the QGP hadronizes~\cite{Thews:2000rj}, measurements of the quarkonium elliptic flow can, in principle, tell us at what time quarkonia are formed by recombination. Therefore it is important to understand how the elliptic flow transmits from heavy quarks to quarkonia. In our study, the momentum distributions of $b$ and $\bar{b}$ are chosen as:
\be
E\frac{\diff^3N}{\diff p^3} =\frac{1}{2\pi} \frac{\diff^2N}{p_T\diff p_T\diff y}\big(1+2v_2^{(b)}\cos(2\phi)\big)\,,
\ee
where $\phi$ is the angle around the $z$-axis. The initial $p_T$ distribution is taken from the FONLL calculation for $2.76$ TeV Pb-Pb collision at rapidity $y=0$~\cite{Cacciari:1998it}. Pairs of $b$ and $\bar{b}$ are sampled and recombined by gluon emission according to the rate at $T=250$ MeV assuming they are at the same position. The $v_2$ of produced $\Upsilon$ is computed by averaging $\cos(2\phi)$ in each $p_T$ bin with size $1$ GeV. The results are plotted in Fig.~\ref{fig:upsilon_v2}. At low $p_T$, the distribution becomes isotropic as expected. As $p_T$ increases, the curves are flatten out. We note that at high $p_T$ fragmentation becomes the dominant mechanism, which will be studied in future work. In the plotted $p_T$ range where recombination dominates, the quarkonium $v_2$ is sensitive to that of heavy quarks.

\begin{figure}
\centering
\includegraphics[width=1\linewidth]{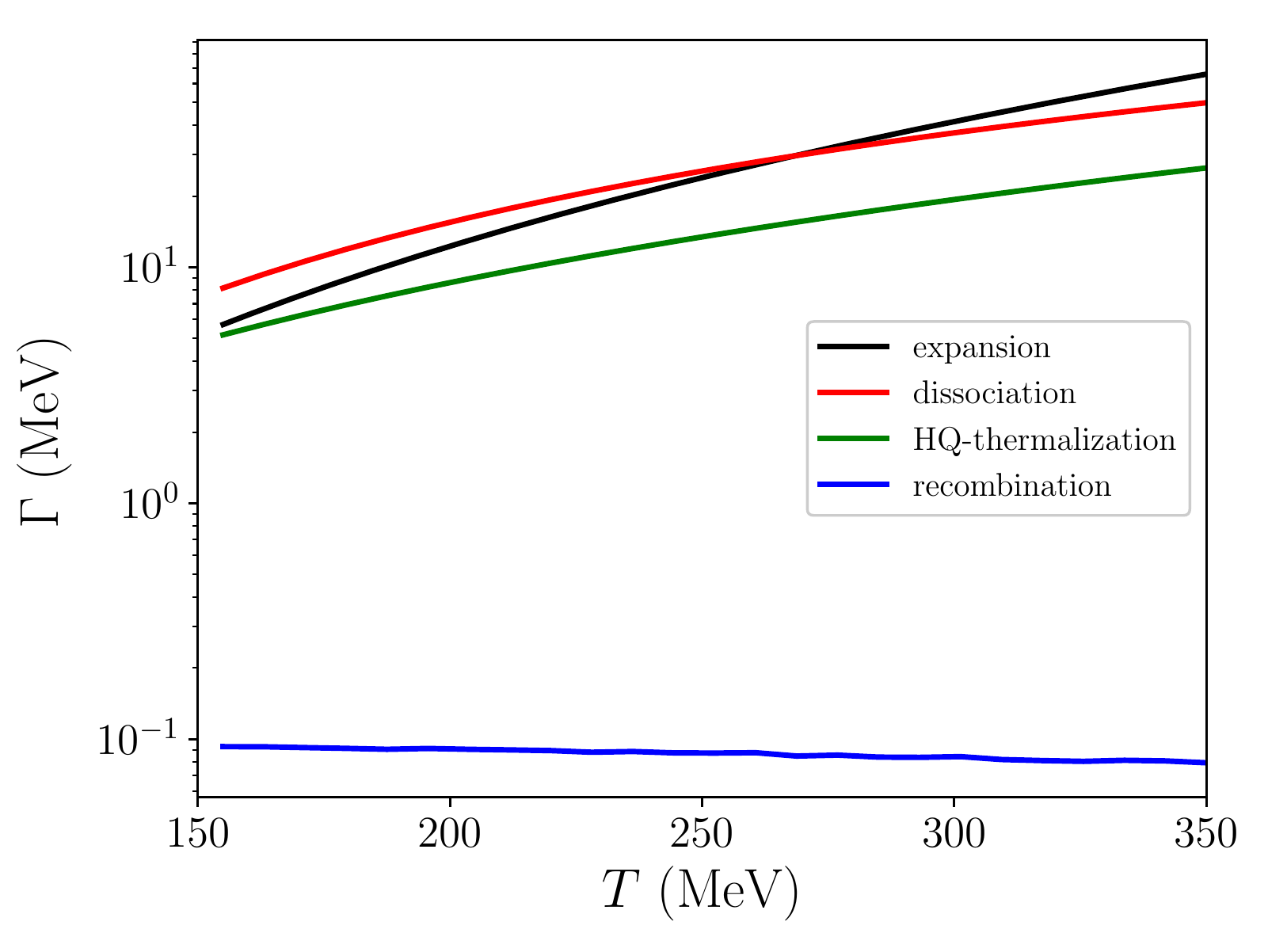}
\caption{(Color online) Thermal rates of expansion, dissociation, HQ thermalization and recombination.}\label{fig:rates}
\end{figure}

\begin{figure}
\centering
\includegraphics[width=1\linewidth]{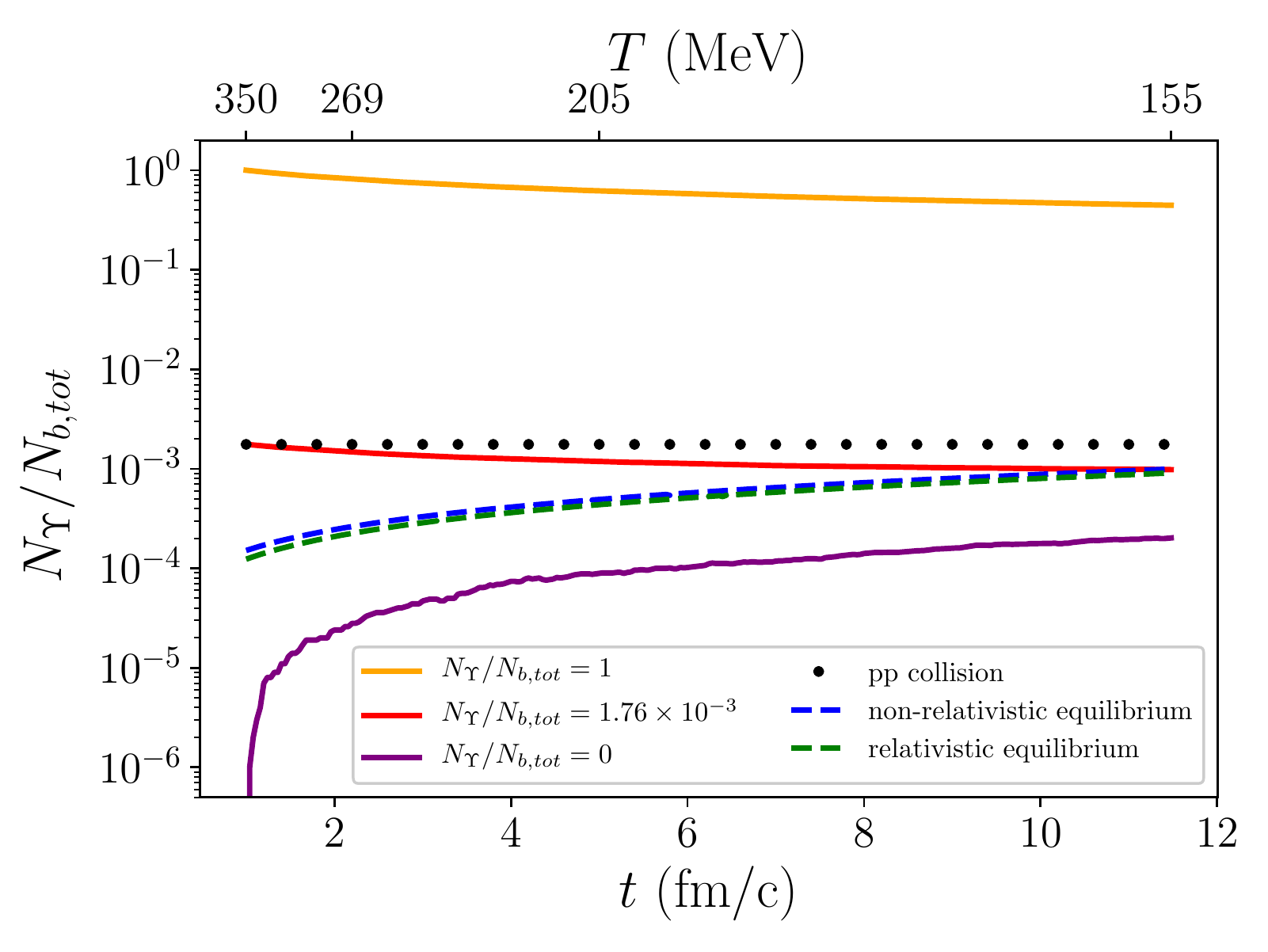}
\caption{(Color online) Evolution of the $\Upsilon(1S)$ fraction in QGP undergoing a boost invariant expansion. The upper (yellow) and lower (purple) solid curves correspond to the cases where all $b,\bar{b}$ quarks are assumed to be initially bound and free respectively. The horizontal dotted line indicates the $\Upsilon(1S)$ fraction measured in pp collisions. The middle (red) solid curve  represents the average of the upper and lower ones weighted in such a way that it starts at the measured pp $\Upsilon(1S)$ fraction.}\label{fig:bjorken}
\end{figure}

Finally we study the dynamics of the system under a boost invariant longitudinal expansion~\cite{Bjorken:1982qr}. The temperature dependence is the Bjorken model given by
\be
T=T_0\Big(\frac{t_0}{t}\Big)^{c_s^2}\,.
\ee
Here we assume $t_0=1$ fm/c, $T_0=350$ MeV and a speed of sound $c_s^2=1/3$. The various rates are plotted as a function of temperature in Fig.~\ref{fig:rates}: the expansion rate defined as $|dT/dt|/T$, the dissociation rate of a static $\Upsilon$, the thermally averaged recombination rate, and the HQ relaxation (thermalization) rate.

We simulate the system starting either at $N_b = 5, N_{\Upsilon} = 0$ or at $N_b = 0, N_{\Upsilon} = 5$ with HQ diffusion. The initial momenta of $b$ and $\bar{b}$ are randomly sampled angularly with the magnitude distributed according to the $p_T$ spectrum in the same FONLL calculation as above. The momentum distribution of $\Upsilon$ is given by the convolution of those of $b$ and $\bar{b}$. The evolution of the $\Upsilon(1S)$ fraction is shown in Fig.~\ref{fig:bjorken}. The fraction in pp collision is roughly $1.76\times10^{-3}$~\cite{Du:2017qkv} and is indicated by the dotted horizontal line. For comparison, we take the weighted average of the two simulations with initial fraction $0$ or $1$ so that the initial fraction starts at the pp value.

For the curve starting at $N_{\Upsilon}/N_{b,\ma{tot}} = 1$ (all $b,\bar{b}$ initially bound) dissociation is the dominant process. Because the dissociation and expansion rates are on the same order, as shown in Fig.~\ref{fig:rates}, the survival probability of $\Upsilon$ is large and the curve stays far away from equilibrium at the end of expansion. 

On the other hand, the curve starting at $N_{\Upsilon}/N_{b,\ma{tot}} = 0$ (all $b,\bar{b}$ initially free) always fall below the equilibrium. The reason is two-fold: The recombination is significantly slower than the expansion as shown in Fig.~\ref{fig:rates} and the thermalization of HQ is not fast enough. We also studied simulations without HQ diffusion and find that the influence of HQ diffusion is small in this scenario. However, if all the rates except the expansion rate were larger, the curve including HQ diffusion would be closer to the equilibrium curve, though it would still lag behind.

Lastly, the curves starting at the pp fraction happen to approach the equilibrium line in the end, but this does not indicate the system reaches equilibrium. 
The recombination contribution here is negligible. However, we note that both the equilibrium fraction and recombination contribution depend on the value of $N_{b,\ma{tot}}$. For much larger values of $N_{b,\ma{tot}}$ the equilibrium and recombination curves would move up and the recombination contribution would be significant, similar to what is observed in the charm sector.

It can be seen that different initial conditions lead to largely distinct final ratios. Since we do not fully understand the initial production of quarkonia in heavy ion collisions, we can hope to learn this together with the in-medium evolution from experiments. The change during the hadronic phase should be small due to the small cross sections~\cite{Lin:2000ke}. Therefore, it is important to measure the final ratios of hidden and open heavy flavors in various $p_T$ and rapidity ranges as a function of centrality and collision energy. We will gain more information from these measurements on the quarkonium production mechanism. Thus, it is essential to do the measurements at both the CERN Large Hadron Collider and BNL Relativistic Heavy Ion Collider.

In summary, we have used pNRQCD to calculate the dissociation and recombination rates of $\Upsilon(1S)$ in a thermal QGP. We studied the dynamics of the $b\bar{b}-\Upsilon$ system in QGP via the Boltzmann transport equation, which we solved by Monte-Carlo simulation. We showed how the system reaches equilibrium starting from different initial conditions. We demonstrated the importance of HQ diffusion in the medium: It is necessary for the system to reach equilibrium. We then calculated the elliptic flow of $\Upsilon$ produced from recombinations. We argued that measurements of $v_2(\Upsilon)$ probe stages of quarkonia production by recombination. Finally we studied the system under a Bjorken expansion. We showed that different initial $\Upsilon$ fractions evolve to widely different final results and argued that measurements on the hidden-to-open heavy flavor ratio could address open questions in quarkonium production in heavy ion collisions. 

We acknowledge helpful discussions with Steffen Bass, Weiyao Ke, Michael Strickland and Yingru Xu. XY thanks Nora Brambilla, Miguel Escobedo, Jacopo Ghiglieri, P\'eter Petreczky and Antonio Vairo for discussions on pNRQCD and acknowledges the hospitality of the nuclear theory group at Brookhaven National Laboratory where part of this work was completed. XY acknowledges support from U.S. Department of Energy (Research Grant No. DE-FG02-05ER41367) and Brookhaven National Laboratory.

\bibliographystyle{apsrev4-1}

\end{document}